\documentclass[10pt,prd,singlecolumn,
nofootinbib,
noshowkeys,
noshowpacs,
superscriptaddress,
floatfix
]{revtex4-2}
\usepackage{amsmath,amsfonts,amsthm,amssymb}
\usepackage[dvips]{graphics,graphicx}
\usepackage[usenames,dvipsnames]{color}
\definecolor{darkblue}{RGB}{0,0,196}
\definecolor{darkgreen}{RGB}{0,120,0}
\usepackage[colorlinks=true,linktocpage=true,linkcolor=darkblue,citecolor=red,urlcolor=darkblue]{hyperref}
\usepackage{cancel}
\usepackage{bbold}
\usepackage{multirow}
\usepackage{longtable}
\usepackage{color}
\usepackage[normalem]{ulem}
\usepackage{hyperref}
\usepackage{bigints}
\usepackage{xparse}
\usepackage{physics}
\usepackage{verbatim}
\usepackage{minibox}
\usepackage{comment}
\usepackage{appendix}
\usepackage{slashed}
\usepackage{marginnote}
\usepackage{graphicx}
\usepackage[nice]{nicefrac}
 
\usepackage{amsmath}
\usepackage[colorinlistoftodos]{todonotes}
\usepackage{hepunits}
\include{commands}
\usepackage{stackengine,scalerel}
\newcommand\hstar[1]{\ThisStyle{\ensurestackMath{%
\setbox0=\hbox{$\SavedStyle#1$}%
\stackengine{0pt}{\copy0}{\kern.2\ht0\smash{\SavedStyle\star}}{O}{c}{F}{T}{S}}}}
\definecolor {darkgreen}{rgb}{0.2,0.7,0.2}

\newcommand{\di}{{\rm d}}

\def\spt{{\cal S}}
\def\wT{{\widehat T}}
\def\wj{{\widehat j}}

\def\wP{{\widehat P}}

\def\wspt{{\widehat{\cal S}}} 

\def\wPhi{{\widehat{\Phi}}}

\def\wrho{{\widehat{\rho}}}
\def\wrhol{{\widehat{\rho}_{\rm LE}}}

\newcommand{\be}{\begin{equation}}
\newcommand{\ee}{\end{equation}}                                                                               
\def\bea{\begin{eqnarray}}
\def\eea{\end{eqnarray}}

\begin{document}
\title{Entropy current and entropy production in relativistic spin hydrodynamics}
\author{Francesco Becattini}
\email{becattini@fi.infn.it}
\affiliation{Department of Physics, University of Florence and INFN
Via G. Sansone 1, I-50019, Sesto F.no (Firenze), Italy}
\author{Asaad Daher}
\email{asaad.daher@ifj.edu.pl}
\affiliation{Institute  of  Nuclear  Physics  Polish  Academy  of  Sciences,  PL-31-342  Krak\'ow,  Poland}
\author{Xin-Li Sheng}
\email{sheng@fi.infn.it}
\affiliation{INFN Sezione di Firenze, Via G. Sansone 1, I-50019, Sesto F.no (Firenze), Italy}
\begin{abstract}
We use a first-principle quantum-statistical method to derive the expression of the 
entropy production rate in relativistic spin hydrodynamics. We show that  
the entropy current is not uniquely defined and can be changed by means of entropy-gauge 
transformations, much the same way as the stress-energy tensor and the spin tensor can
be changed with pseudo-gauge transformations. We show that the local thermodynamic relations,
which are admittedly educated guesses in relativistic spin hydrodynamics inspired by
those at global thermodynamic equilibrium, do not hold in general and they are also non-invariant 
under entropy-gauge transformations. Notwithstanding, we show that the entropy production 
rate is independent of those transformations and we provide a universally applicable expression,
extending that known in literature, from which one can infer the dissipative parts of the 
energy momentum and spin tensors.
\end{abstract}
\maketitle

\section{Introduction}

Motivated by the evidence of spin polarization of particles produced in relativistic heavy 
ion collisions~\cite{STAR:2017ckg,Niida:2018hfw}, there is a growing interest in the so-called
relativistic spin hydrodynamics~\cite{Becattini:2011zz,Florkowski:2017ruc,Florkowski:2018fap,Montenegro:2017rbu,Hattori:2019lfp,
Fukushima:2020ucl,Daher:2022xon,She:2021lhe,Hongo:2021ona,Gallegos:2021bzp,Gallegos:2022jow,Peng:2021ago,
Cao:2022aku,Weickgenannt:2022zxs,Weickgenannt:2022qvh,Biswas:2023qsw,Xie:2023gbo}. 
Relativistic spin hydrodynamics 
stipulates that the description of a relativistic fluid requires the addition of a {\em spin tensor},
that is the mean value of a rank 3 tensor operator $\wspt^{\lambda\mu\nu}$ (the last two
indices anti-symmetric) contributing to the overall angular momentum current:
$$
 \widehat{\cal J}^{\lambda\mu\nu} = x^\mu \wT^{\lambda\nu} - x^\nu \wT^{\lambda \mu}
 + \wspt^{\lambda\mu\nu}\,,
$$
where $\wT^{\mu\nu}$ is the stress-energy tensor operator.
This current is conserved, which implies that the spin tensor fulfills the continuity equation:
\be
\partial_{\lambda}\wspt^{\lambda\mu\nu}=\wT^{\nu\mu}-\wT^{\mu\nu}\,.
\ee

It is important to point out that the spin tensor - and the stress-energy tensor as well - are not 
uniquely defined. Indeed, they can be changed with a so-called pseudo-gauge transformation
\cite{Hehl:1976vr,Becattini:2011ev,Speranza:2020ilk} to a new couple of tensors fulfilling the same dynamical equations 
and providing the same integrated conserved charges. Since the spin tensor can be made vanishing 
with a suitable pseudo-gauge transformation, its dynamical meaning has been questioned, yet it was 
observed in ref.~\cite{Becattini:2018duy} that for a fluid {\em not} in global thermodynamic equilibrium 
(such as the QGP throughout its lifetime) the quantum state of the system (i.e. the density operator 
describing initial local equilibrium) is not invariant under pseudo-gauge transformations. Thus, 
in principle, the physical measurements depend on the pseudo-gauge if the initial quantum state is 
not invariant, and particularly on the intensive quantity which is thermodynamically conjugated to 
the spin tensor, the spin potential. For instance, if the spin tensor does not vanish, the spin polarization 
of final particles depends on the difference between spin potential and thermal vorticity \cite{Buzzegoli:2021wlg}.
The microscopic conditions underpinning spin hydrodynamics have been studied and elucidated in 
ref.~\cite{Hongo:2021ona}, where it was made clear that spin hydrodynamics regime occurs under 
a specific hierarchy of the interaction time scales in the system.

A key problem in the relativistic hydrodynamics with spin is the derivation of the constitutive equations
of the spin tensor as well as of the anti-symmetric part of the stress-energy tensor. This problem has drawn 
significant attention over the past few years, with several derivations of constitutive equations ~\cite{Hattori:2019lfp,Fukushima:2020ucl,Daher:2022xon,She:2021lhe,Hongo:2021ona,Gallegos:2021bzp,Gallegos:2022jow} 
based on the requirement of the positivity of local entropy production rate. 
However, like in the traditional approach to relativistic hydrodynamics, the entropy current is not 
really derived, but it is obtained from an educated guess of a thermodynamic relations between the proper 
densities of entropy, energy, charge and ``spin density" $S^{\mu\nu}\equiv u_{\lambda} \spt^{\lambda,\mu\nu}$ 
as follows:
\be\label{tradtherm}
\begin{split}
    & Ts + \mu n = \rho + p - \frac{1}{2} \omega_{\mu\nu} S^{\mu\nu} \\
    & \di p = s \, \di T + n \, \di \mu +\frac{1}{2} S^{\mu\nu} \di \omega_{\mu\nu} 
\end{split}
\ee
where $T$ is temperature, $\mu$ a chemical potential, $\rho$ the proper energy density, $p$ the
pressure, $n$ the charge density and $\omega_{\mu\nu}$ is the spin potential\footnote{This is related 
to $\Omega$ defined in ref.~\cite{Becattini:2018duy} by the relation $\omega = T \Omega$.}.

In this work, we apply the quantum-statistical approach to relativistic hydrodynamics
~\cite{ChGvanWeert, Zubarev, Becattini:2014yxa} by including spin tensor. The quantum statistical 
method based on local equilibrium density operator has several advantages over other approaches
in that it makes it possible to {\em derive} from first principles a form of the entropy current 
and entropy production rate rather than constructing it assuming a particular form of the local thermodynamic 
relation such as the equation \eqref{tradtherm}. We will use a recent result on the extensivity of the logarithm 
of the partition function to obtain an exact form of the entropy current \cite{Becattini:2019poj}.
We will be able to show that the relation \eqref{tradtherm} is incomplete and that the entropy density has,
in general, additional terms involving the spin tensor. Furthermore, we will extend the derivation of
ref. \cite{ChGvanWeert} of the entropy production rate to include the spin tensor. Such general 
relation is the starting point to derive the constitutive relations for the anti-symmetric part
of the stress-energy tensor and the spin tensor.

\section{Entropy current and local equilibrium}
\label{sec2}

In the quantum-statistical description of a relativistic fluid, the local equilibrium density operator denoted 
as $\wrho_{LE}$ is obtained by maximizing entropy $S = - \Tr (\wrho \log \wrho)$ over some preset 
space-like hypersurface by constraining the mean values of the energy, momentum, charge and spin densities 
to be equal to their actual values~\cite{Becattini:2018duy}:
\begin{align}\label{densityoperator}
    \wrho_{\rm LE}=\frac{1}{Z_{\rm LE}}\exp\left[-\int_{\Sigma} \di\,\Sigma_{\mu}\,\left(\wT^{\mu\nu}\beta_{\nu}
    -\zeta \wj^\mu - \frac{1}{2}\Omega_{\lambda\nu}\wspt^{\mu\lambda\nu}\right)\right]\,,
\end{align}
where $\di \Sigma_\mu \equiv \di \Sigma \, n_\mu$, $n$ being the unit vector perpendicular to the 
hypersurface $\Sigma$; the function $Z_{\rm LE}$ is the partition function, and the operators $\wT^{\mu\nu}$, 
$\wspt^{\mu\lambda\nu}$ are the energy-momentum and spin tensor operators, a particular couple amongst all 
the possible couples connected by pseudo-gauge transformations. The constraints read:
\be\label{constraints}
 n_\mu T^{\mu\nu} = n_\mu T^{\mu\nu}_{\rm LE}\,, \qquad n_\mu j^{\mu} = n_\mu j^\mu_{\rm LE}\,,
 \qquad  n_\mu \spt^{\mu\lambda\nu} = n_\mu \spt^{\mu\lambda\nu}_{\rm LE}\,,
\ee
where the local equilibrium values are defined as:
\be\label{leqmean}
 X_{\rm LE} \equiv \Tr (\wrho_{\rm LE} \widehat X) - \bra{0} \widehat X \ket{0}\,,
\ee
with $\ket{0}$  being the supposedly non-degenerate lowest lying eigenvector of the operator in the exponent
of \eqref{densityoperator}. In the equation \eqref{densityoperator}, the fields $\beta_{\nu}$, $\zeta$ 
and $\Omega_{\lambda\nu}$ are the Lagrange multipliers related to this problem, and they are the 
thermal velocity four-vector, the chemical potential to temperature ratio, and the spin potential to temperature 
ratio respectively, that is:
\be\label{lagrange}
 \beta = \frac{u}{T}\,,  \qquad  \zeta = \frac{\mu}{T}\,, \qquad \Omega = \frac{\omega}{T}\,.
\ee
It is worth pointing out that they can be obtained as solutions of the constraint equations \eqref{constraints} 
\cite{Becattini:2014yxa}, if the exact values of the stress-energy tensor and other currents is known. 
In relativistic hydrodynamics, since they are not known {\em a priori}, they are solutions 
of the hydrodynamic partial differential equations with initial conditions expressed by the equations 
\eqref{constraints} over the initial Cauchy space-like hypersurface. It should also be stressed that $\beta$ 
thereby defines a so-called hydrodynamic frame in its own (the so-called thermodynamic or thermometric or 
$\beta$ frame), which does not coincide with the Landau or Eckart frames. At global equilibrium one has:
\be\label{equilibrium}
 \beta_\mu = b_\mu + \varpi_{\mu\nu} x^\nu\,, \;\; {\rm with} \;\; b,\varpi = {\rm const}\,, 
  \qquad \Omega = \varpi\,, \qquad \zeta = {\rm const}\,,
\ee
where $\varpi$ is a constant anti-symmetric tensor, the thermal vorticity. 

Starting from the equation \eqref{densityoperator}, it is possible to prove \cite{Becattini:2019poj} 
that if the operator:
$$
\widehat \Upsilon \equiv \int_{\Sigma} \di\,\Sigma_{\mu}\,\left(\wT^{\mu\nu}\beta_{\nu}
    -\zeta \wj^\mu - \frac{1}{2}\Omega_{\lambda\nu}\wspt^{\mu\lambda\nu} \right)
$$
is bounded from below and the lowest lying eigenvalue $\ket{0}$ is non-degenerate, the 
logarithm of $Z_{\rm LE}$ is {\em extensive}, namely it can be written as an integral over $\Sigma$:
\be\label{extensive}
\log Z_{\rm LE} = \int_\Sigma \di \Sigma_\mu \; \phi^\mu 
- \bra{0} \widehat \Upsilon \ket{0} = \int_\Sigma \di \Sigma_\mu \; \left[ \phi^\mu 
- \bra{0}\left(\wT^{\mu\nu}\beta_{\nu} -\zeta \wj^\mu - \frac{1}{2}\Omega_{\lambda\nu}\wspt^{\mu\lambda\nu} 
\right)\ket{0} \right]
\ee
where
\be\label{phi}
 \phi^{\mu}=\int_{1}^{\infty} \di\lambda \; \left( T^{\mu\nu}_{\rm LE}(\lambda) \beta_{\nu}
 -\zeta j^\mu_{\rm LE}(\lambda) - \frac{1}{2}\Omega_{\lambda\nu} \spt^{\mu\lambda\nu}_{\rm LE}(\lambda)
 \right) 
\ee
is defined as the thermodynamic potential current. In the equation \eqref{phi}, the integration 
variable $\lambda$ is a dimensionless parameter which multiplies the exponent of the local equilibrium 
density operator \eqref{densityoperator}, that is:
\be\label{rholambda}
    \wrho_{\rm LE}(\lambda) =\frac{1}{Z_{\rm LE}(\lambda)}
    \exp\left[-\lambda \int_{\Sigma} \di\,\Sigma_{\mu}\,\left(\wT^{\mu\nu}\beta_{\nu}
    -\zeta \wj^\mu - \frac{1}{2}\Omega_{\lambda\nu}\wspt^{\mu\lambda\nu}\right)\right]\,,
\ee
and $T^{\mu\nu}_{\rm LE}(\lambda), j^\mu_{\rm LE}(\lambda), \spt^{\mu\lambda\nu}_{\rm LE}(\lambda)$ 
are calculated with the equation \eqref{leqmean} with the modified density operator just defined 
in the eq.~\eqref{rholambda}. 
As $\lambda$ multiplies $\beta$, $\zeta$ and $\Omega$, this coefficient plays the role of a rescaled inverse 
temperature, so it possible to change the integration variable in \eqref{phi} from $\lambda$ to 
$T'(x) = T(x)/\lambda$ and rewrite the thermodynamic potential current:
\be\label{phit}
 \phi^{\mu}(x) =\int_{0}^{T(x)} \frac{\di T'}{T^{\prime 2}} \; \left( T^{\mu\nu}_{\rm LE}(x)[T',\mu,\omega] 
  u_{\nu}(x) -\mu(x) j^\mu_{\rm LE}(x)[T',\mu,\omega] - \frac{1}{2} \omega_{\lambda\nu}(x) 
  \spt^{\mu\lambda\nu}_{\rm LE}(x)[T',\mu,\omega] \right) \,,
\ee
where we used the eq.~\eqref{lagrange}. The equation \eqref{phit} shows that the thermodynamic 
potential current can be calculated by integrating in temperature 
the mean values at local thermodynamic equilibrium of the various involved currents. It is important to
stress the meaning of the square brackets, which denote a {\em functional} dependence on the arguments.
Indeed, the local equilibrium values of the currents at some point $x$ depend not just on the value
of $T',\mu,\omega$ at the same point $x$, but on the whole functions $T'(y),\mu(y),\omega(y)$; 
tantamount, assuming analiticity, on the value of the functions and all their gradients at the point $x$.

Once the thermodynamic potential current $\phi^\mu$ is determined, an entropy current can be defined.
By using the definition \eqref{leqmean} and the equations \eqref{densityoperator},\eqref{extensive} we 
have:
\be\label{totalent}
\begin{split}
S =& -\Tr (\wrho_{\rm LE} \log \wrho_{\rm LE}) = \log Z_{\rm LE} +  \int_\Sigma \di \Sigma_\mu \; 
\left( \Tr(\wrhol \wT^{\mu\nu}) \beta_{\nu} -\zeta \Tr( \wrhol \wj^\mu) -
  \frac{1}{2}\Omega_{\lambda\nu}\Tr( \wrhol \wspt^{\mu\lambda\nu}) \right) \\
  =& \int_\Sigma \di \Sigma_\mu \;\left(  \phi^\mu + T^{\mu\nu}_{\rm LE}\beta_{\nu} 
  -\zeta j^\mu_{\rm LE} -\frac{1}{2} \Omega_{\lambda\nu}\spt^{\mu\lambda\nu}_{\rm LE} \right)\,,
\end{split}
\ee
which implies that we can define an entropy current as:
\be\label{def1}
 s^\mu = \phi^\mu + T^{\mu\nu}_{\rm LE}\beta_{\nu} -\zeta j^\mu_{\rm LE} -
  \frac{1}{2}\Omega_{\lambda\nu}\spt^{\mu\lambda\nu}_{\rm LE}\,.
\ee
%

\section{Entropy current: quasi-objective form and entropy-gauge transformations}
\label{sec3}

The equations \eqref{phit} and \eqref{def1} define the fields $\phi^\mu$ and $s^\mu$. However, they
depend not just on $x$ but also on the space-like hypersurface employed to define the local equilibrium mean 
values of the currents through the density operator \eqref{densityoperator} $\wrhol$. More specifically, 
to each point $x$ there must be a corresponding hypersurface $\Sigma$ needed to define local
thermodynamic equilibrium through the constraints \eqref{constraints}. Altogether, to define the 
thermodynamic potential and entropy current at each point $x$ one needs to specify in advance a family 
of 3D space-like hypersurfaces, a so-called {\em foliation} of the space-time. 
\begin{figure}[h]
\begin{center}
 \includegraphics[width=9cm]{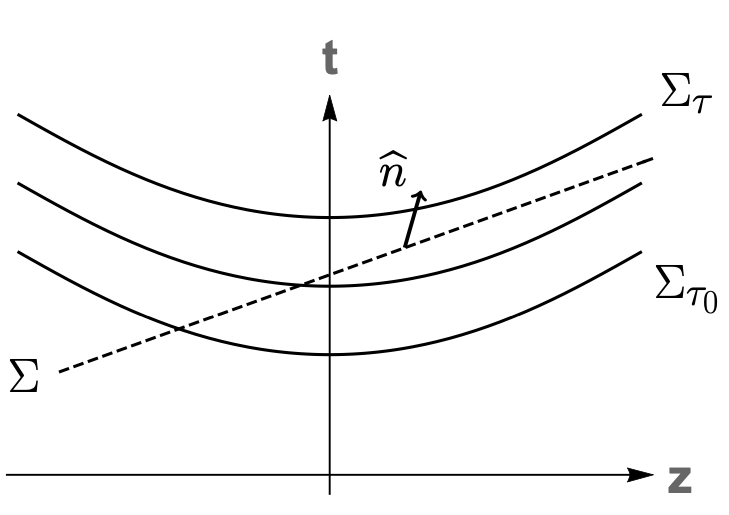}
 \caption{An example of a family of 3D space-like hypersurfaces (solid lines) defining a foliation 
 parametrized by the real variable $\tau$, which is necessary to define local thermodynamic 
 equilibrium. The 3D space-like hypersurface $\Sigma$ does not belong to the foliation.}
 \label{foliationfig1}
\end{center}
\end{figure}

The dependence of the currents \eqref{phit} and \eqref{def1} on the foliation involves a problem in 
that, if we are to calculate the total entropy by integrating the entropy current \eqref{def1} on some 
$\Sigma$ which does not belong to the foliation (see figure~\ref{foliationfig1}), the result is in general 
different from the total entropy which would be obtained from the Von Neumann formula imposing the 
constraints of local equilibrium \eqref{constraints} over this particular $\Sigma$. In symbols:
\be\label{sproblem}
  \int_\Sigma \di\Sigma_\mu \; s^\mu \ne - \Tr(\wrhol(\Sigma) \log \wrhol(\Sigma))\,,
\ee
with equality applying only if $\Sigma$ belongs to the foliation. Such a situation is quite disturbing,
as one of the requested features of the entropy current field is to provide the actual value of the total 
entropy.

To settle the issue, one can define the entropy current more in general by using the actual values
of the conserved currents instead of their values at local equilibrium, that is:
\be\label{def2}
 s^\mu = \phi^\mu + T^{\mu\nu} \beta_{\nu} -\zeta j^\mu - \frac{1}{2}\Omega_{\lambda\nu} \spt^{\mu\lambda\nu}\,,
\ee
with, accordingly (omitting most arguments to make the expression more compact):
\be\label{phi2}
 \phi^{\mu}=\int_{0}^{T} \frac{\di T'}{T^{\prime 2}} \; \left( T^{\mu\nu}[T'] u_{\nu}
 -\mu j^\mu[T'] - \frac{1}{2} \omega_{\lambda\nu} \spt^{\mu\lambda\nu}[T']
 \right) \, .
\ee
Indeed, by using the actual values of the currents, whenever we integrate them over some 
hypersurface $\Sigma$, not necessarily belonging to the original foliation, the result is the same 
that we would have obtained by enforcing the constraints \eqref{constraints} on $\Sigma$ itself. Since:
\begin{align*}
\int_\Sigma \di \Sigma_\mu s^\mu & = \int_\Sigma \di \Sigma \; n_\mu \left( \phi^\mu + T^{\mu\nu} \beta_{\nu} 
-\zeta j^\mu - \frac{1}{2}\Omega_{\lambda\nu} \spt^{\mu\lambda\nu} \right) \\
& = \int_\Sigma \di \Sigma \; n_\mu \left( \int_{0}^{T} \frac{\di T'}{T^{\prime 2}} \; \left( T^{\mu\nu}[T'] 
u_{\nu} -\mu j^\mu[T'] - \frac{1}{2} \omega_{\lambda\nu} \spt^{\mu\lambda\nu}[T'] \right) + 
T^{\mu\nu} \beta_{\nu} - \zeta j^\mu - \frac{1}{2}\Omega_{\lambda\nu} \spt^{\mu\lambda\nu} \right)
\end{align*}
a local equilibrium density operator like in equation \eqref{densityoperator} can be built on 
$\Sigma$ by enforcing the constraints \eqref{constraints} therein. Therefore, the above expression 
becomes, by using the eqs. \eqref{phi} and \eqref{totalent}:
\begin{align*}
& \int_\Sigma \di \Sigma \; n_\mu \left( \int_{0}^{T} \frac{\di T'}{T^{\prime 2}} \; 
\left( T^{\mu\nu}_{\rm LE}[T'] u_{\nu} -\mu j^\mu_{\rm LE}[T'] - \frac{1}{2} \omega_{\lambda\nu} 
\spt^{\mu\lambda\nu}_{\rm LE}[T']\right)  + T^{\mu\nu}_{\rm LE} \beta_{\nu} 
- \zeta j^\mu_{\rm LE} - \frac{1}{2}\Omega_{\lambda\nu} \spt^{\mu\lambda\nu}_{\rm LE} \right) \\
& = - \Tr(\wrhol(\Sigma) \log \wrhol(\Sigma))\,.
\end{align*}
Similarly, the equation \eqref{extensive} can be extended to a relation which applies
to any space-like hypersurface $\Sigma$:
\be\label{extext}
\int \di \Sigma_\mu \phi^\mu =\log Z_{\rm LE}(\Sigma) + \bra{0} \widehat \Upsilon \ket{0}
\ee

The two equations \eqref{def2} and \eqref{phi2} are the final expressions defining the entropy current 
for a system which is close to local thermodynamic equilibrium. It is worth remarking that those equations 
imply that the entropy current depends on the actual mean values of the conserved currents ensuing from 
the quantum field Lagrangian. The question is whether, with the definitions \eqref{phi2} and \eqref{def2} 
the thermodynamic potential and entropy current fields are {\em objective}, namely independent of a predefined 
foliation. For this purpose, in the first place the Lagrange multiplier fields $\beta,\zeta,\Omega$ 
which also appear in those definitions should be independent thereof, a condition which is achieved 
in relativistic hydrodynamics if they are obtained as solutions of partial differential equations from 
given initial conditions. 

Yet, a complete independence cannot be achieved. Looking carefully at the thermodynamic potential current 
in the eq. \eqref{phi2}, it appears that its definition involves the knowledge of the conserved currents 
as functionals of the temperature. However, such functionals can be constructed only if the local 
equilibrium operator is introduced, hence a separation between the local equilibrium term and 
the dissipative term, 
which does require the introduction of a foliation. We can signify this limitation by saying that the 
thermodynamic potential current, and the entropy current as well, can be made {\em quasi-objective}. 
The quasi-objective nature of the entropy current also shows up in the entropy production rate
\eqref{entrate}, as will be discussed in Section \ref{sec5}.

A further issue is that the thermodynamic potential current and the entropy current fields
are not unique. It is quite clear that a transformation of the thermodynamic potential current:
\be\label{entgauge}
  \phi^{\prime \mu} = \phi^\mu + \nabla_\lambda A^{\lambda\mu}\,,
\ee
where $A^{\lambda\mu}$ is an arbitrary anti-symmetric tensor, implying:
\be\label{entgauge2}
s^{\prime\mu} = s^\mu + \nabla_\lambda A^{\lambda\mu}
\ee
will leave the total entropy
$$
   S = \int_\Sigma \di \Sigma_\mu \; s^\mu
$$
invariant because of the relativistic Stokes theorem, provided that the tensor $A$ fulfills suitable 
boundary conditions. Therefore, just like $T^{\mu\nu}$ and $\spt^{\mu\lambda\nu}$, the entropy 
current $s^\mu$ is not uniquely defined and can be changed with transformations \eqref{entgauge2}
\cite{Li:2020eon} \footnote{Specific transformations of the entropy current were introduced in
ref.~\cite{Li:2020eon} in the context of pseudo-gauge transformations of the stress-energy and spin
tensor, see Section~\ref{sec6}.}, henceforth defined as {\em entropy-gauge transformations}. Such 
a freedom in defining the entropy current affects the local thermodynamic relations, as we will see.
Nevertheless, the divergence of the entropy current is invariant under pseudo-gauge transformations
because:
$$
\nabla_\mu s^{\prime \mu} = \nabla_\mu s^\mu + \nabla_\mu \nabla_\lambda A^{\lambda\mu}
= \nabla_\mu s^\mu\,.
$$
The entropy production rate will be discussed in Section \ref{sec5}.

\section{Discussion on local thermodynamic relations}
\label{sec4}

The local thermodynamic relation between proper densities can be obtained by contracting the entropy 
current with a suitable four-velocity vector. For instance, one can contract the \eqref{def2} with the 
four-velocity defined by the direction of $\beta$ that is $u^\mu = \beta^\mu/\sqrt{\beta^2} = T \beta^\mu$
\footnote{Note that if one contracts the \eqref{def2} with the normalized time-like eigenvector of 
the stress-energy tensor $u_L$, which defines with the Landau frame, the obtained local thermodynamic
relations reads:
\be\label{ltr2}
 s_L = s^\mu u_{L\mu} = \phi \cdot u_L + u_L \cdot \beta \rho_L - \zeta n_L - \frac{1}{2} \Omega_{\lambda\nu} 
 u_{L\mu} \spt^{\mu\lambda\nu} \equiv \phi \cdot u_L + u_L \cdot \beta \rho_L - \zeta n_L - 
 \frac{1}{2} \Omega_{\lambda\nu} S^{\lambda\nu}_L
\ee
Since $u_L \cdot \beta \ne \sqrt{\beta^2} = 1/T$, it turns out that, even if the entropy current was 
quasi-objective, the local thermodynamic relation is frame-dependent \cite{Becattini:2014yxa} and much 
care should be taken when using it to derive constitutive equations.}:
\be\label{ltr1}
  s \equiv s^\mu u_\mu = \phi \cdot u + \frac{1}{T} \rho - \zeta n - \frac{1}{2} \Omega_{\lambda\nu} 
  u_\mu \spt^{\mu\lambda\nu}
  \equiv \phi \cdot u + \frac{1}{T} \rho - \frac{\mu}{T} n - \frac{1}{2} \Omega_{\lambda\nu} S^{\lambda\nu}\,,
\ee
where $\rho=u_\mu u_\nu T^{\mu\nu}$ and $n=u_\mu j^\mu$. Defining the pressure as:
$$
  p \equiv T \phi \cdot u\,,
$$
the eq.~\eqref{ltr1} coincides with the first thermodynamic relation in eq.~\eqref{tradtherm}. It should
be pointed out though, that only at global equilibrium with $\beta = {\rm const}$ this quantity coincides with
the hydrostatic pressure, that is the diagonal spatial component of the mean value of the stress-energy tensor
(see Appendix \ref{app1}); in all other cases, it does not need to. By contracting the eq.~\eqref{phi2}
with $u = \beta/\sqrt{\beta^2}$ we obtain:
\be
\begin{split} \label{thermodynamic_pressure}
 p = T \phi \cdot u &= T \int_{0}^{T} \frac{\di T'}{T^{\prime 2}} \; \left( u_\mu T^{\mu\nu}[T'] u_\nu
 -\mu u_\mu j^\mu[T'] - \frac{1}{2} \omega_{\lambda\nu} u_\mu \spt^{\mu\lambda\nu}[T'] 
  \right)
 \\
 & =  T \int_{0}^{T} \frac{\di T'}{T^{\prime 2}} \; \left( \rho[T'] -\mu n[T'] 
 - \frac{1}{2} \omega_{\lambda\nu} S^{\lambda\nu}[T'] \right)\,,
\end{split}
\ee
whence the following relation can be readily obtained:
\be\label{pderiv}
\frac{\partial p}{\partial T}\Big|_{\mu,\omega} = s
\ee
by using the \eqref{ltr1}. This equation is the first step in proving the second relation \eqref{tradtherm},
but in fact the remaining two partial derivative of the pressure function do not need to coincide with the
charge density and the spin density and in general:
$$
\frac{\partial p}{\partial \mu}\Big|_{T,\omega} \ne n\,, \qquad \qquad 
\frac{\partial p}{\partial \omega_{\lambda\nu}}\Big|_{T,\mu} \ne S^{\lambda\nu}\,.
$$
Indeed, for the equality to apply, one would need the following thermodynamic relation to hold:
\be
 T \di s = \di \rho -\mu \, \di n - \frac{1}{2} \omega_{\lambda\nu} \di S^{\lambda\nu}\,,
\ee
and yet this cannot be obtained from the definitions \eqref{ltr1} and \eqref{phi2}.

Furthermore, the relation \eqref{pderiv} is not invariant under entropy-gauge transformations. The 
thermodynamic potential current can be redefined according to the \eqref{entgauge} and, contracting
with the four-velocity we get:
$$
 p' = T \phi'\cdot u = p + T u_\mu \nabla_\lambda A^{\lambda\mu}\,, 
$$
where the transformed quantities are denoted with a prime. It is then easy to show that:
\be\label{noninvar}
  \frac{\partial p'}{\partial T}\Big|_{\mu\omega} = s' +  u_\mu
  T\frac{\partial}{\partial T} \nabla_\lambda A^{\lambda\mu}\Big|_{\mu\omega} \,.
\ee
If the second term on the right hand side is non-vanishing, even the relation \eqref{pderiv} is
broken. An example of an entropy-gauge transformation which breaks the \eqref{pderiv} is shown
in Appendix \ref{app3} for the global equilibrium with rotation.

In conclusion, the local thermodynamic relations \eqref{tradtherm} are not fully appropriate 
in the derivation of the divergence of the entropy current. On one 
hand, it turns out that the differential relation in \eqref{tradtherm} cannot be proved in general 
and on the other hand, perhaps most importantly, because they are both non-invariant under entropy-gauge 
transformations, even for the case of a global equilibrium with $\varpi \ne 0$.

\section{Entropy production rate}
\label{sec5}

The entropy production rate, which is important to obtain the constitutive equations of relativistic
hydrodynamics, is determined by taking the divergence of the equation \eqref{def2}. By 
using the continuity equations of the stress-energy tensor, the number current and the spin tensor, 
that is:
\be\label{contin}
 \nabla_\mu T^{\mu\nu} = 0\,, \qquad \nabla_\mu j^\mu = 0\,, \qquad \nabla_{\mu}\spt^{\mu\lambda\nu}
 =T^{\nu\lambda}-T^{\lambda\nu}\,,
\ee
we obtain:
\begin{align}\label{entropyproduction}
\begin{split}
\nabla_{\mu}s^{\mu} & =\nabla_{\mu}\phi^{\mu}+T^{\mu\nu}\nabla_{\mu}\beta_{\nu}-j^\mu \nabla_\mu \zeta 
-\frac{1}{2} \spt^{\mu\lambda\nu}\nabla_{\mu} \Omega_{\lambda\nu}- \frac{1}{2}\Omega_{\lambda\nu} 
\nabla_{\mu}\spt^{\mu\lambda\nu} \\
& = \nabla_{\mu}\phi^{\mu}+T_S^{\mu\nu}\xi_{\mu\nu} -j^\mu \nabla_\mu \zeta + 
 T_A^{\mu\nu} (\Omega_{\mu\nu} - \varpi_{\mu\nu})
-\frac{1}{2} \spt^{\mu\lambda\nu}\nabla_{\mu} \Omega_{\lambda\nu}\,,
\end{split}
\end{align}
where $T_S$ and $T_A$ are the symmetric and anti-symmetric parts of the stress-energy tensor and
$$
  \xi_{\mu\nu} = \frac{1}{2} \left( \nabla_{\mu}\beta_{\nu}+ \nabla_{\nu}\beta_{\mu} \right)\,,
  \qquad \varpi_{\mu\nu} = \frac{1}{2} \left( \nabla_{\nu}\beta_{\mu} - \nabla_{\mu}\beta_{\nu} \right)
$$
are the thermal shear and thermal vorticity tensor respectively. 
\begin{figure}[h]
\begin{center}
 \includegraphics[width=9cm]{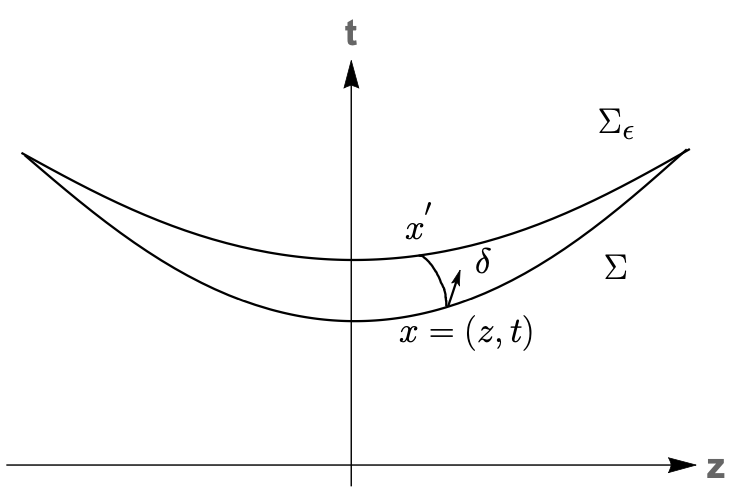}
 \caption{The hypersurface $\Sigma$ is mapped to $\Sigma_{\epsilon}$
 by an infinitesimal diffeomorphism $\epsilon$. The vector field $\delta(x)$ is defined as 
 $\di x'(x,\epsilon)/\di \epsilon|_{\epsilon=0}$.}
 \label{Diffeomorphic hypersurfaces}
\end{center}
\end{figure}

The next step, as it appears from the equation \eqref{entropyproduction}, is the calculation of the 
divergence of the thermodynamic potential current, $\nabla_{\mu}\phi^{\mu}$. To derive it, it is 
convenient to study the change of $\log Z_{\rm LE}$ under an infinitesimal change of the integration 
3D hypersurface (see figure~\ref{Diffeomorphic hypersurfaces}). An infinitesimal change of hypersurface may be seen, 
in simple terms, as the result of locally moving every point $x \in \Sigma$ to a point $x'(x,\epsilon) \in 
\Sigma_\epsilon$, $\epsilon$ being a finite real parameter. Setting $x'(x,0) = x$, we define:
$$
\frac{\di x'^\mu(x,\epsilon)}{\di \epsilon}\Big|_{\epsilon=0} = \delta^\mu(x)\,.
$$
For a small $\epsilon$, the vector field $\delta$ loosely represents the direction in which the 
hypersurface is locally modified and the parameter $\epsilon$ describes how far along the vector field 
$\delta$ we move the hypersurface. Formally, these definitions are those of a one-parameter group of
diffeomorphisms, which are a prerequisite to define the Lie derivative. For the special case of the 
integration of a vector field $V^\mu$ over a 3D-hypersurface, one has (see Appendix~\ref{app2}):
\begin{align}\label{dsigma}
\lim_{\epsilon\to\ 0} \frac{1}{\epsilon} \left( \int_{\Sigma_\epsilon} \di \Sigma_{\mu}V^{\mu}-
\int_{\Sigma} \di \Sigma_{\mu} \, V^{\mu} \right) = \int_{\partial \Sigma} \di \tilde S_{\mu\nu}\, 
\delta^{\mu} V^{\nu} +\int_{\Sigma} \di \Sigma \cdot \delta \, \nabla_{\mu} V^{\mu}\,,
\end{align}
where $\partial \Sigma$ is the 2-D boundary surface. We can apply this equation to the \eqref{extext}
to obtain the infinitesimal change of $\log Z_{\rm LE}$ by a change of the hypersurface:
\begin{align}\label{liederiv1}
& \lim_{\epsilon\to\ 0} \frac{1}{\epsilon} \left[ \log Z_{\rm LE}(\Sigma_\epsilon) - \log Z_{\rm LE}(\Sigma) \right] 
 = \int_{\Sigma} \di \Sigma \cdot \delta \, \nabla_{\mu} \left( \phi^{\mu} - \bra{0} \wT^{\mu\nu} \beta_{\nu} - 
 \zeta \wj^\mu - \frac{1}{2}\Omega_{\lambda\nu} \wspt^{\mu\lambda\nu} \ket{0} \right) \nonumber \\
 & =  \int_{\Sigma} \di \Sigma \cdot \delta \, \nabla_{\mu} \phi^{\mu} - 
  \int_{\Sigma} \di \Sigma \cdot \delta \, \bra{0} \wT_S^{\mu\nu} \xi_{\mu\nu} - 
   \wj^\mu \nabla_\mu \zeta + \wT_A^{\mu\nu} (\Omega_{\mu\nu} - \varpi_{\mu\nu}) - 
  \frac{1}{2} \wspt^{\mu\lambda\nu} \nabla_\mu \Omega_{\lambda\nu} \ket{0}\,,
\end{align}
where, in the last step, we have used the continuity equations \eqref{contin}, holding at operator level.
On the other hand, the logarithm of the partition function can be calculated by means of its definition 
as a trace. For an infinitesimal $\epsilon$ one has:
\begin{align*}
& Z_{\rm LE}(\Sigma_\epsilon) = \Tr \left( \exp\left[ - \int_{\Sigma_\epsilon}\di \Sigma_{\mu}
\left(\wT^{\mu\nu}\beta_{\nu}- \zeta \wj^\mu - \frac{1}{2}\Omega_{\lambda\nu}\wspt^{\mu\lambda\nu}\right)\right]\right)
\nonumber\\
&\simeq \Tr\left( \exp\left[ - \int_{\Sigma} \di \Sigma_{\mu}\left(\wT^{\mu\nu}\beta_{\nu} 
- \zeta \wj^\mu - \frac{1}{2}\Omega_{\lambda\nu}\wspt^{\mu\lambda\nu}\right) - 
 \epsilon \int_{\Sigma} \di \Sigma \cdot \delta
\nabla_{\mu}\left(\wT^{\mu\nu}\beta_{\nu} -\zeta \wj^\mu - \frac{1}{2}\Omega_{\lambda\nu}\wspt^{\mu\lambda\nu}
\right)\right] \right) \\
& = \Tr\left( \exp\left[ - \int_{\Sigma} \di \Sigma_{\mu} \left( \wT^{\mu\nu}\beta_{\nu} 
- \zeta \wj^\mu - \frac{1}{2}\Omega_{\lambda\nu}\wspt^{\mu\lambda\nu} \right) \right. \right. \\
& - \left. \left. \epsilon \int_{\Sigma} \di \Sigma \cdot \delta \; \left( \wT_S^{\mu\nu}\xi_{\mu\nu} -\wj^\mu \nabla_\mu \zeta + 
 \wT_A^{\mu\nu} (\Omega_{\mu\nu} - \varpi_{\mu\nu}) - \frac{1}{2} \wspt^{\mu\lambda\nu}\nabla_{\mu} \Omega_{\lambda\nu} 
 \right) \right] \right)\,,
\end{align*}
where we have used the equation \eqref{dsigma} - assuming that the boundary term vanishes - and, again,
the continuity equations \eqref{contin} at operator level. By expanding the trace in the small parameter 
$\epsilon$, and keeping in mind the equation \eqref{densityoperator}, we obtain:
\begin{align*}
 Z_{\rm LE}(\Sigma_\epsilon) & \simeq  Z_{\rm LE}(\Sigma) - \epsilon Z_{\rm LE}(\Sigma) \\
 &\times \int_{\Sigma} \di \Sigma \cdot \delta \; \left( \Tr(\wrhol \wT_S^{\mu\nu}) \xi_{\mu\nu} - 
   \Tr(\wrhol \wj^\mu) \nabla_\mu \zeta + \Tr(\wrhol \wT_A^{\mu\nu}) (\Omega_{\mu\nu} - \varpi_{\mu\nu}) - 
  \frac{1}{2} \Tr(\wrhol \wspt^{\mu\lambda\nu}) \nabla_\mu \Omega_{\lambda\nu} \right)\,,
\end{align*}
whence:
\begin{align}\label{liederiv2}
& \lim_{\epsilon\to\ 0} \frac{1}{\epsilon} \left[ \log Z_{\rm LE}(\Sigma_\epsilon) - \log Z_{\rm LE}(\Sigma) \right] \\
\nonumber
& = - \int_{\Sigma} \di \Sigma \cdot \delta \; \left( \Tr(\wrhol \wT_S^{\mu\nu}) \xi_{\mu\nu} -
   \Tr(\wrhol \wj^\mu) \nabla_\mu \zeta + \Tr(\wrhol \wT_A^{\mu\nu}) (\Omega_{\mu\nu} - \varpi_{\mu\nu}) - 
  \frac{1}{2} \Tr(\wrhol \wspt^{\mu\lambda\nu}) \nabla_\mu \Omega_{\lambda\nu} \right)\,. 
\end{align}
Therefore, by comparing the equation~\eqref{liederiv1} with the equation~\eqref{liederiv2}, taking
into account that both $\Sigma$ and the field $\delta$ are arbitrary, we can infer that: 
\begin{align}\label{phidiverg}
\nabla_\mu \phi^{\mu} &= - \left[ \left( 
  \Tr(\wrhol \wT_S^{\mu\nu}) -\bra{0} \wT_S^{\mu\nu} \ket{0} \right) \xi_{\mu\nu} - 
 \left( \Tr(\wrhol \wj^\mu) -  \bra{0} \wj^{\mu} \ket{0} \right) \nabla_\mu \zeta \right. \nonumber \\
& + \left. \left( \Tr(\wrhol \wT_A^{\mu\nu}) - \bra{0} \wT_A^{\mu\nu} \ket{0} \right) (\Omega_{\mu\nu} - \varpi_{\mu\nu}) 
 - \frac{1}{2} \left( \Tr(\wrhol \wspt^{\mu\lambda\nu}) - \bra{0} \wspt^{\mu,\lambda\nu} \ket{0} \right)
 \nabla_\mu \Omega_{\lambda\nu} \right] \nonumber \\
 &=  - \left( T^{\mu\nu}_{S\rm (LE)} \xi_{\mu\nu} - 
 j^{\mu}_{\rm LE} \nabla_\mu \zeta + T^{\mu\nu}_{A \rm (LE)} (\Omega_{\mu\nu} - \varpi_{\mu\nu}) 
 - \frac{1}{2} \spt^{\mu\lambda\nu}_{\rm LE} \nabla_\mu \Omega_{\lambda\nu} \right)\,,
\end{align}
where, in the last step, we have used the definition of local equilibrium values.

Now, substituting back the eq.~\eqref{phidiverg} into the eq.~\eqref{entropyproduction}, we obtain the 
evolution of entropy current:
\be\label{entrate}
\nabla_{\mu}s^{\mu}=\left( T_S^{\mu\nu}-T^{\mu\nu}_{S(\rm LE)}\right) \xi_{\mu\nu}
- \left( j^\mu-j^\mu_{\rm LE}\right) \nabla_\mu \zeta  + 
\left( T^{\mu\nu}_{A}-T^{\mu\nu}_{A(\rm LE)}\right) (\Omega_{\mu\nu}-\varpi_{\mu\nu}) 
- \frac{1}{2}\left( \spt^{\mu\lambda\nu}-\spt^{\mu\lambda\nu}_{\rm LE} \right) \nabla_{\mu}\Omega_{\lambda\nu} \,.
\ee
The equation \eqref{entrate} is the main result of this work and it is the starting point to
derive the constitutive equations of dissipative spin hydrodynamics, which relate the anti-symmetric
part of the stress-energy tensor and the spin tensor to the gradients of the spin potential and
the difference between spin potential and thermal vorticity, besides the (thermal) shear tensor
and the gradient of $\zeta=\mu/T$. In the above form, it is in fact a generalization of the one found 
by Van Weert and Zubarev~\cite{ChGvanWeert,Zubarev}, with the addition of the last two terms involving
spin tensor and the spin potential. We stress that the formula \eqref{entrate} is exact and not an 
approximation at some order of a gradient expansion. Indeed, with respect to all previous assessments 
of dissipative spin hydrodynamics based on different approaches 
~\cite{Hattori:2019lfp,Fukushima:2020ucl,Daher:2022xon,She:2021lhe,Hongo:2021ona,Gallegos:2021bzp,Gallegos:2022jow},
a novel feature is apparently the simultaneous appearance of the last two terms of the right hand side. 
While the last term is neglected in almost all derivations, it was actually obtained in ref.~\cite{She:2021lhe}. 
However, it should be pointed out that some terms in previous derivations may have been omitted 
because of a gradient power counting method. A complete analysis 
of the constitutive equations implied by the \eqref{entrate} will be presented in a forthcoming study.

\section{Entropy Current and pseudo-gauge transformations}
\label{sec6}

A question that may arise at this point is whether the change in the entropy current \eqref{def2}
induced by a so-called pseudo-gauge transformations of the stress-energy and spin tensor (see the 
discussion in the Introduction section):
\begin{align}\label{pseudogauge}
    &\wT^{\prime \mu\nu}=\wT^{\mu\nu}+\frac{1}{2}\nabla_{\lambda}\left(\wPhi^{\lambda\mu\nu}-
    \wPhi^{\mu\lambda\nu}-\wPhi^{\nu\lambda\mu}\right),\\
    &\wspt^{\prime \mu\lambda\nu}=\wspt^{\mu\lambda\nu}-\wPhi^{\mu\lambda\nu}\nonumber. 
\end{align}
where $\wPhi^{\mu\lambda\nu}$ is an arbitrary rank-3 tensor anti-symmetric operator in the last two indices,
comes down to an entropy-gauge transformation~\eqref{entgauge2}. If this was the case, the entropy 
production rate would be invariant under a pseudo-gauge transformation. 

By plugging the equations~\eqref{pseudogauge} in~\eqref{def2}~\eqref{phi2} and with the simplifying
assumption $\Omega=\varpi$ we get (see Appendix \ref{app4} for the full derivation):
\begin{align}
 \phi^{\prime \mu}&=\phi^{\mu} +\int_{0}^{T}\frac{\di T^\prime}{T^\prime} \; 
 \left[\nabla_{\lambda}A^{\lambda\mu}-\Phi^{\lambda\mu\nu} \xi_{\lambda\nu}\right],\label{phipg}\\
 s^{\prime\mu}&= s^{\mu} + \int_{0}^{T}\frac{\di T^\prime}{T^\prime} \; 
 \left[\nabla_{\lambda}A^{\lambda\mu}-\Phi^{\lambda\mu\nu} \xi_{\lambda\nu}\right]
 +\nabla_{\lambda}A^{\lambda\mu}-\Phi^{\lambda\mu\nu}\xi_{\lambda\nu}\label{spg}.
\end{align}
where $A^{\lambda\mu}=(1/2) \beta_{\nu}\left(\Phi^{\lambda\mu\nu}-\Phi^{\nu\lambda\mu}+\Phi^{\mu\nu\lambda}\right)$ 
is an anti-symmetric rank 2 tensor and $\xi_{\lambda\nu}$ is the thermal shear tensor:
$$
\xi_{\lambda\nu} = \frac{1}{2} \left( \nabla_\lambda \beta_\nu + \nabla_\nu \beta_\lambda \right)
$$
The last term on the right-hand side of equations~\eqref{phipg},\eqref{spg} cannot be written 
as a total derivative of an anti-symmetric tensor. In essence, the equations \eqref{phipg} and 
\eqref{spg} show that a general pseudo-gauge transformation of stress-energy and spin tensor 
\eqref{pseudogauge} does not lead to an entropy-gauge transformation. Therefore, the divergence 
of the entropy current is not invariant under a pseudo-gauge transformation and a simultaneous
entropy-gauge transformation cannot be used to restore the relation:
$$
 s^\mu = \phi^\mu + T^{\mu\nu} \beta_\nu - \zeta j^\mu - \frac{1}{2} 
  \varpi_{\lambda\nu}\spt^{\mu\lambda\nu}   \; ;
$$
for the new quantities; in this respect, our conclusion differs from that of ref.~\cite{Li:2020eon}.

This conclusion holds provided that the intensive thermodynamics fields $\beta$ and $\zeta$ 
(and, in the most general case $\Omega$) are left unchanged by the pseudo-gauge transformations. 
In virtually all known non-equilibrium cases, including the one discussed in the Section~\ref{sec2}, 
their definition is based on stress-energy and spin tensors, so they also get changed 
under pseudo-gauge transformations. Nevertheless, at global equilibrium with rotation, or, more in
general, with $\varpi \ne 0$, $\Omega=\varpi$ and $\beta$, $\zeta$ are invariant as $\beta$ 
is a Killing vector and $\zeta$ a constant.

\section{Discussion and conclusions}
\label{discuss}

The formula \eqref{entrate} shows that entropy production rate, in general, is non-vanishing whenever there
is a difference between the actual value of the conserved (or conserved-related) currents and the corresponding
values at local thermodynamic equilibrium, such as $T_{S(\rm LE)}$, $j_{\rm LE}$, etc. 
As we have emphasized in this paper, local equilibrium depends on the choice of a family of 3D space-like 
hypersurfaces, i.e. a foliation. In relativistic hydrodynamics, this freedom ultimately corresponds to the
choice of a four-velocity vector, so-called hydrodynamic frame. The dependence on the foliation shows up
in the divergence of the entropy current \eqref{entrate}, which is manifestly dependent on local equilibrium 
values (see the discussion at the end of Section \ref{sec3}). 

We emphasize that the formula \eqref{entrate} is exact, not an approximation at some order of a gradient 
expansion. In other words, fixing the order in a gradient expansion of hydrodynamic quantities is 
not required to obtain it. However, for future work, once constitutive equations are determined, a gradient 
ordering can be made based on the involved scales in the physical problem.

In conclusion, in this work we have employed a quantum-statistical approach to derive the entropy 
current and entropy production rate without assuming the traditional local thermodynamic relations~\eqref{tradtherm}. 
In fact, we have shown that the local thermodynamic relations do not hold in general and that they are also
non-invariant under allowed transformations of the entropy current, that we have defined as entropy-gauge
transformations. We have obtained an expression of the entropy production rate \eqref{entrate} which 
extends to spin hydrodynamics previous expression obtained in refs.~\cite{ChGvanWeert,Zubarev}. This form
is especially well-suited to derive the constitutive equations of dissipative spin hydrodynamics, what
will be the subject of a forthcoming work.

\medskip

\section*{Acknowledgements}

Part of this work was carried out in the workshop "The many faces of relativistic fluid dynamics" 
held in the Kavli institute in Santa Barbara (CA) USA, supported in part by the National Science 
Foundation under Grants No. NSF PHY-1748958 and PHY-2309135. F.B. gratefully acknowledges fruitful discussions with the participants in the workshop, especially J. Armas, G. Denicol, P. Kovtun and 
M. Hippert Teixeira.
Interesting discussions with R. Ryblewski, E. Grossi and A. Giachino are also
acknowledged. A.D. thanks the Department of Physics, University of Florence and INFN for the hospitality. A.D. acknowledges the financial support provided by the Polish National Agency for Academic Exchange NAWA  
under the Programme STER–Internationalisation of doctoral schools, Project no.  PPI/STE/2020/1/00020 
and the Polish National Science Centre Grants No.2018/30/E/ST2/00432.

\appendix

\section{Thermodynamic potential current at homogeneous global equilibrium}
\label{app1}

Homogeneous global equilibrium is defined by the condition $\beta=\mbox{const}$ i.e. vanishing thermal vorticity 
in the equation \eqref{equilibrium}. Plugging this form in the equation \eqref{densityoperator}, the density 
operator takes on the familiar form (for simplicity, we assume there are no charges in the system):
\begin{align}
\wrho_{\rm GE}=\frac{1}{Z}\exp\left[- \beta \cdot \wP\right]\,.
\end{align}
Due to the symmetries of the above operator, the mean value of the stress-energy tensor operator has the 
ideal form:
\be\label{setgeq}
 T^{\mu\nu} = \Tr (\wrho_{\rm GE} \wT^{\mu\nu}) - \bra{0} \wT^{\mu\nu} \ket{0} = 
  (\rho + p) u^\mu u^\nu - p g^{\mu\nu}\,,
\ee
where $u \equiv \beta/{\sqrt{\beta^2}}$.
According to eq.~\eqref{phi2}, the thermodynamic potential current is:
\begin{align}\label{phiint}
  \phi^{\mu} = \int_{0}^{T} \frac{\di T'}{T^{\prime 2}} \; T^{\mu\nu}[T'] u_{\nu} 
  = \int_{0}^{T} \frac{\di T'}{T^{\prime 2}} \; \rho[T'] u^\mu \,,
\end{align}
where $T = 1/\sqrt{\beta^2}$.
The above expression confirms the expectation that, at the homogeneous global equilibrium,
any vector field should be parallel to $\beta$ with a coefficient depending on $\beta^2$ 
or, equivalently, the temperature $T$. Therefore:
\be\label{phiform}
\phi^{\mu}=\beta^{\mu}\phi(\beta^2)
\ee
and the goal is now to show that such scalar coefficient $\phi(\beta^2)$ is just
the pressure, as defined by the equation \eqref{setgeq}.

By taking the derivative with respect to $\beta$ of the partition function, we have:
\be\label{logzder}
-\frac{\partial \log Z}{\partial\beta_{\nu}}= \Tr(\wrho_{\rm GE} \wP^{\nu}) = 
 \int_\Sigma \di \Sigma_\mu \; \Tr(\wrho_{\rm GE} \wT^{\mu\nu})\,.
\ee
Since:
$$
 \log Z = \int \di \Sigma_\mu \; \phi^\mu - \beta_\nu \bra{0} \wP^\nu \ket{0} 
$$
from the \eqref{extensive}, from the \eqref{logzder} we can obtain the following equality
\begin{align}
 \int_\Sigma \di \Sigma_{\mu}\left(\frac{\partial\phi^{\mu}}{\partial \beta_{\nu}}\right)=
 - \int_\Sigma \di \Sigma_{\mu} T^{\mu\nu}\,,
\end{align}
where we have used the \eqref{setgeq}. Since the integration hypersurface $\Sigma$ is 
arbitrary, being at global equilibrium, we can infer the following relation:
$$
   T^{\mu\nu}=-\frac{\partial \phi^{\mu}}{\partial\beta_{\nu}}+
  \partial_{\lambda}A^{\lambda\mu\nu}\,,
$$
where the rank 3 tensor $A^{\lambda\mu\nu}$ is anti-symmetric in the indices $\lambda\mu$. 
Such a gradient
term is allowed by the Stokes theorem in Minkowski space-time if suitable boundary conditions are 
fulfilled. Yet, since the equilibrium is homogeneous, it must vanish due to traslational invariance as
all mean values ought to be constant and uniform. This implies that, by using the equation \eqref{phiform}:
\begin{align}\label{A7}
  T^{\mu\nu} =-\frac{\partial \phi^{\mu}}{\partial\beta_{\nu}} = -\frac{\partial}{\partial\beta_{\nu}} 
  \phi \beta^\mu
 = -\phi g^{\mu\nu}+T\frac{\partial \phi}{\partial T} u^\mu u^\nu\,.
\end{align}
We can now compare~\eqref{setgeq} with~\eqref{A7} and infer that $\phi=p$ and consequently 
$\phi^{\mu}=p \beta^{\mu}$. 
By plugging the latter equation in the \eqref{phiint} and taking the derivative with respect to $T$ 
we obtain:
$$
 T \frac{\partial p}{\partial T} = \rho + p\,,
$$
which makes also the second term on the right hand side of equation \eqref{A7} consistent with the identification
$\phi=p$.

\section{Lie derivatives and integration}
\label{app2}

Suppose we have a one-parameter group of diffeomorphisms $x'(x,\epsilon)$ with $\epsilon$ a real number.
Let $\omega$ be a rank 3 differential form which is to be integrated over a 3D hypersurface embedded in 
the 4D space-time. We denote with $\omega'_\epsilon$ the differential form which is obtained from 
$\omega$ through the diffeomorphism, that is:
$$
  \omega'_{\epsilon}(x)_{\mu_1\mu_2\mu_3} = J_{\mu_1}^{\nu_1} J_{\mu_2}^{\nu_2}J_{\mu_3}^{\nu_3} \;
  \omega(x'(x,\epsilon))_{\nu_1,\nu_2,\nu_3}
$$
where $J_\mu^\nu = \partial x'(x,\epsilon)^\nu/\partial x^\mu$ is the jacobian matrix element of the 
diffeomorphism.
Let $\Sigma_\epsilon$ be the image of the hypersurface $\Sigma$ through the diffeomorphism. Then we have:
$$
\int_{\Sigma_\epsilon} \omega = \int_{\Sigma} \omega'_\epsilon 
$$
whence:
\begin{align*}
\lim_{\epsilon \to 0} \frac{1}{\epsilon} \left( \int_{\Sigma_\epsilon} \omega(x) -  
\int_{\Sigma} \omega(x) \right) = 
\lim_{\epsilon \to 0} \frac{1}{\epsilon} \int_{\Sigma} \omega'_\epsilon(x) - \omega(x)
= \int_{\Sigma} {\cal L}_\delta (\omega(x))\,,
\end{align*}
where ${\cal L}_\delta$ stands for the Lie derivative along the vector field $\delta(x) = 
\di x'(x,\epsilon)/\di \epsilon|_{\epsilon=0}$.

The so-called Cartan magic formula can now be used in the last expression, leading to:
\be\label{cartan}
 \int_{\Sigma} {\cal L}_\delta (\omega) = \int_{\Sigma} i_\delta \di \omega + \di (i_\delta \omega)
 =  \int_{\Sigma} i_\delta \di \omega + \int_{\partial \Sigma} i_\delta \omega\,,
\ee
where $i_\delta$ stands for the interior product of the form with the vector field $\delta$ and 
$\di$ stands for the exterior derivative. The second term on the right hand side of \eqref{cartan}
is an integral of an exterior derivative and it has been turned into a 2D boundary integral of 
$i_\delta \omega$ by using the generalized Stokes theorem for differential forms. 

We can apply the above formulae to the differential form which is the dual of a vector field
$V$ in a 4D space-time, namely:
\be\label{omegadual}
    \omega_{\mu\nu\rho} = \frac{1}{6} E_{\mu\nu\rho\sigma} V^\sigma =
    \frac{1}{6} \sqrt{|g|} \epsilon_{\mu\nu\rho\sigma} V^\sigma\,.
\ee
With this form, it can be shown that:
\be\label{intdual}
\int_{\Sigma} \omega(x) =  \int_{\Sigma} \di \Sigma_\mu V^\mu\,.
\ee
The exterior derivative can be readily worked out by using the definition:
$$
 (\di \omega)_{\lambda\mu\nu\rho} = - \frac{1}{24} E_{\lambda\mu\nu\rho} \nabla \cdot V\,,
$$
which leads, by using the definition of interior product, to:
$$
 (i_\delta \di \omega)_{\mu\nu\rho} = \frac{1}{6} E_{\mu\nu\rho\sigma} \delta^\sigma \nabla \cdot V\,.
$$
Therefore, by using the above expression and the \eqref{intdual} we get:
$$
\int_{\Sigma}  i_\delta \di \omega = \int_{\Sigma} \di \Sigma_\mu \, \delta^\mu \nabla \cdot V\,.
$$
The second integral in the \eqref{cartan} can be similarly worked out and one eventually 
obtains the equation \eqref{dsigma}.

\section{Non-invariance of the local thermodynamic relations: an example}
\label{app3}

We are going to show that the local thermodynamic relation \eqref{pderiv} is not invariant under
entropy-gauge transformation \eqref{entgauge}, namely that the equation \eqref{noninvar} applies
with a non-trivial second term in the right hand side.

We consider, as specific example, global equilibrium with non-vanishing thermal vorticity in the 
equation \eqref{equilibrium}. Let:
$$
   A^{\lambda\mu} = f(\kappa^2) \varpi^{\lambda\mu}\,,
$$
where $\kappa^\mu = \varpi^{\mu\nu} \beta_\nu$ and $f(\kappa^2)=g(\kappa^2)/\kappa^2$ with $g(\kappa^2)$
an adimensional differentiable function (this form of $f(\kappa^2)$ ensures that $A^{\lambda\mu}$ has the
correct dimension for the entropy gauge transformation \eqref{entgauge}). One has, in Cartesian coordinates:
$$
\nabla_\lambda A^{\lambda\mu} = \partial_\lambda A^{\lambda\mu} = f'(\kappa^2) 
 \varpi^{\lambda\mu} \partial_\lambda \kappa^2 = f'(\kappa^2) \varpi^{\lambda\mu} 2 \kappa^\nu \partial_\lambda 
 \kappa_\nu
  = f'(\kappa^2) \varpi^{\lambda\mu} 2 \kappa^\nu \partial_\lambda (\varpi_{\nu\rho} \beta^\rho)
  = f'(\kappa^2) \varpi^{\lambda\mu} 2 \kappa^\nu \varpi_{\nu\rho} \varpi^\rho_{\; \lambda}\,,
$$
where, in the last step, we have used the relation $\varpi_{\mu\nu} = \partial_\nu \beta_\mu$ which applies at global
equilibrium where $\partial_\mu \beta_\nu + \partial_\nu \beta_\mu = 0$.

Now let $\gamma^\rho = \varpi^{\rho\nu} \kappa_\nu$ so that:
\be\label{firststep}
 \partial_\lambda A^{\lambda\mu} = - 2 f'(\kappa^2)  \gamma^\rho \varpi_{\rho\lambda} \varpi^{\lambda\mu} \,.
\ee
Contracting the equation \eqref{firststep} with $u_\mu$ we get:
\be\label{secondstep}
 u_\mu \partial_\lambda A^{\lambda\mu} = T \beta_\mu \partial_\lambda A^{\lambda\mu} =
 - 2 f'(\kappa^2) T \gamma^\rho \varpi_{\rho\lambda} \varpi^{\lambda\mu} \beta_\mu =
 - 2 f'(\kappa^2) T \gamma^\rho \varpi_{\rho\lambda} \kappa^\lambda = - 2 f'(\kappa^2) T \gamma^2\,.
\ee
The derivative in \eqref{noninvar} must be taken by keeping $\omega = T \varpi$ constant. Therefore,
being:
$$
  \kappa^\mu = \varpi^{\mu\nu} \beta_\nu = \frac{1}{T^2} \omega^{\mu\nu} u_\nu\,, 
  \qquad \qquad \gamma^\rho = \varpi^{\rho\nu} \kappa_\nu = \frac{1}{T^3} \omega^{\rho\nu}
 \omega_{\nu\alpha} u^{\alpha} \,,
$$
and choosing $g(\kappa^2)=1$, we have that the expression in the equation \eqref{secondstep} is
proportional to $T^{3}$,
\be\label{secondterm}
  T \frac{\partial}{\partial T} \left( u_\mu \partial_\lambda A^{\lambda\mu} \right)
  = T \frac{2}{(\kappa^2)^2} \gamma^2\,,
\ee
which is non-vanishing. Therefore, using the \eqref{secondterm} in the equation \eqref{noninvar}
we get:
$$
  \frac{\partial p'}{\partial T}\Big|_{\mu\omega} = s' + T \frac{2}{(\kappa^2)^2} \gamma^2\,,
$$
which proves the non-invariance of the local thermodynamic relation.

\section{Pseudo-gauge transformation of the thermodynamic potential and the entropy current}
\label{app4}

We study here the effect of a pseudo-gauge transformation \eqref{pseudogauge} on the 
$$
   T^{\mu\nu} \beta_\nu - \frac{1}{2} \varpi_{\lambda\nu} \spt^{\mu\lambda\nu}
$$
i.e. a part of the entropy current \eqref{def2} with $\Omega_{\lambda\nu}=\varpi_{\lambda\nu}
=\frac{1}{2}(\partial_{\nu}\beta_{\lambda}-\partial_{\lambda}\beta_{\nu})$. By plugging the 
\eqref{pseudogauge} for the mean values, we get:
\be\label{D1}
   T^{\prime\mu\nu} \beta_\nu - \frac{1}{2} \varpi_{\lambda\nu} \spt^{\prime\mu\lambda\nu}
   = T^{\mu\nu} \beta_\nu - \frac{1}{2} \varpi_{\lambda\nu} \spt^{\mu\lambda\nu} 
   + \frac{1}{2}\nabla_{\lambda}(\Phi^{\lambda\mu\nu}-\Phi^{\mu\lambda\nu}-\Phi^{\nu\lambda\mu})
   \beta_{\nu}+\frac{1}{2}\varpi_{\lambda\nu}\Phi^{\mu\lambda\nu}
\ee
The last two terms can be transformed as follows:
\begin{align}\label{D2}
& \frac{1}{2}\nabla_{\lambda}(\Phi^{\lambda\mu\nu}-\Phi^{\mu\lambda\nu}-\Phi^{\nu\lambda\mu})\beta_{\nu}
 +\frac{1}{2}\varpi_{\lambda\nu}\Phi^{\mu\lambda\nu}\nonumber \\
 = & \frac{1}{2}\nabla_{\lambda}(\Phi^{\lambda\mu\nu}-\Phi^{\mu\lambda\nu}-\Phi^{\nu\lambda\mu})\beta_{\nu}
 +\frac{1}{2}\nabla_{\nu}\beta_{\lambda}\Phi^{\mu\lambda\nu},\nonumber\\
= &\frac{1}{2}\nabla_{\lambda}(\Phi^{\lambda\mu\nu}-\Phi^{\mu\lambda\nu}-\Phi^{\nu\lambda\mu})\beta_{\nu}
+\frac{1}{2}\nabla_{\nu}(\beta_{\lambda}\Phi^{\mu\lambda\nu})-\frac{1}{2}\beta_{\lambda}\nabla_{\nu}\Phi^{\mu\lambda\nu}
\nonumber \\
= & \frac{1}{2}(\nabla_{\lambda}\Phi^{\lambda\mu\nu}-\nabla_{\lambda}\Phi^{\nu\lambda\mu})\beta_{\nu}
+\frac{1}{2}\nabla_{\lambda}(\beta_{\nu}\Phi^{\mu\nu\lambda}).
\end{align}
where we have used the definition of thermal vorticity, the Leibniz rule, the antisymmetry of
$\Phi^{\mu\nu\lambda}$ in the last two indices, and, in the last step, the saturated indices 
$\lambda$-$\nu$ of the last two terms have been swapped. Using again the same methods, the 
eq.~\eqref{D2} can be turned into:
\begin{align}\label{D4}
& \frac{1}{2}(\nabla_{\lambda}\Phi^{\lambda\mu\nu}-\nabla_{\lambda}\Phi^{\nu\lambda\mu})\beta_{\nu}
+\frac{1}{2}\nabla_{\lambda}(\beta_{\nu}\Phi^{\mu\nu\lambda}) \nonumber \\
= & \frac{1}{2}\nabla_{\lambda}\left(\beta_{\nu}\Phi^{\lambda\mu\nu}-\Phi^{\nu\lambda\mu}\beta_{\nu}+\beta_{\nu}\Phi^{\mu\nu\lambda}\right)-\frac{1}{2}\Phi^{\lambda\mu\nu}\nabla_{\lambda}\beta_{\nu}+\frac{1}{2}\Phi^{\nu\lambda\mu}
\nabla_{\lambda}\beta_{\nu}\nonumber\\
= & \frac{1}{2}\nabla_{\lambda}\left(\beta_{\nu}\Phi^{\lambda\mu\nu}-\Phi^{\nu\lambda\mu}\beta_{\nu}+\beta_{\nu}\Phi^{\mu\nu\lambda}\right)-\frac{1}{2}\Phi^{\lambda\mu\nu}
\nabla_{\lambda}\beta_{\nu}-\frac{1}{2}\Phi^{\lambda\mu\nu}\nabla_{\nu}\beta_{\lambda} \nonumber \\
= & \frac{1}{2}\nabla_{\lambda}\left(\beta_{\nu}\Phi^{\lambda\mu\nu}-\Phi^{\nu\lambda\mu}\beta_{\nu}+\beta_{\nu}\Phi^{\mu\nu\lambda}\right)-\Phi^{\lambda\mu\nu}\xi_{\lambda\nu}
\end{align}
Defining the anti-symmetric tensor:
$$
A^{\lambda\mu}=\frac{1}{2}\left(\beta_{\nu}\Phi^{\lambda\mu\nu}-\Phi^{\nu\lambda\mu}\beta_{\nu}+\beta_{\nu}\Phi^{\mu\nu\lambda}\right)
$$
the equations \eqref{D1}, \eqref{D2} and \eqref{D4} imply:
\be\label{D5}
 T^{\prime\mu\nu}\beta_\nu - \frac{1}{2}\varpi_{\lambda\nu}\spt^{\prime\mu\lambda\nu}
= T^{\mu\nu} \beta_\nu - \frac{1}{2} \varpi_{\lambda\nu} \spt^{\mu\lambda\nu}  + \nabla_\lambda A^{\lambda\mu}
 -\Phi^{\lambda\mu\nu}\xi_{\lambda\nu}
\ee
By using the definitions \eqref{phi2} and \eqref{def2} with $\Omega=\varpi$ and the above 
equation~\eqref{D5}, it is straightforward to obtain the transformed thermodynamic potential 
current and entropy current in eqs.~\eqref{phipg} and \eqref{spg}.


\end{document}